\documentclass{elsart}

\usepackage{graphicx}
\usepackage{graphics}
\usepackage{amssymb}
\usepackage{amsmath}
\usepackage{bm}
\begin{document}

\begin{frontmatter}

\title{
Thorotrast and in vivo thorium dioxide: numerical simulation of 30 years 
of $\alpha$ radiation absorption by the tissues near a large 
compact source. 
} 

\author{A.Bianconi}.  
\ead{andrea.bianconi@ing.unibs.it}

\address{
Dipartimento di Ingegneria dell'Informazione, \\
Universit\`a degli Studi di Brescia, Via Branze 38, 
and \\
Istituto Nazionale di Fisica Nucleare, Gruppo di Brescia, 
via Valotti 9, \\ 
I-25123 Brescia, Italy}


\begin{abstract}

\noindent 
Background: 

The epidemiology of the slightly radioactive contrast agent 
named Thorotrast presents a very long 
latency period between the injection and the development of 
the related pathologies. It is an  
example of the more general problem posed by a radioactive internal 
contaminant whose effects are not noteworthy in the short 
term but become dramatic in the long period. 
A point that is still to be explored is 
fluctuations (in space and time) in the localized absorption of 
radiation by the tissues. 

\noindent 
Methods: 

A Monte Carlo simulation code has been developed to study over a 30-year 
period the daily absorption of $\alpha$ radiation 
by $\mu$m-sized portions of 
tissue placed at a distance of 0-100 $\mu$m from a model source,  
that approximates a compact 
thorium dioxide source in liver or spleen whose size is $\gtrsim$  
20 $\mu$m. The biological depletion of the daughter nuclei of the thorium 
series is 
taken into account. The initial condition 
assumes chemically purified natural thorium. 

\noindent 
Results: 

Most of the absorbed dose is concentrated in a 25-$\mu$m 
thick layer of tissue, adjacent to the source 
boundary. 
Fluctuations where a target region with a volume of 1 $\mu$m$^3$ 
is hit by 3-5 $\alpha$ 
particles in a day or in a shorter period of time 
are relevant in a 1-10 $\mu$m thick layer of tissue  
adjacent to the source boundary, where their  
frequency is larger than the Poisson-law 
prediction. 
\end{abstract} 


\end{frontmatter}

\maketitle

\section{Introduction}

Thorotrast epidemiology is an important source of information 
about the long-term risks of exposure to small 
amounts of radioactivity from a thorium source inside the body.  

Thorotrast was used on thousands of patients as a contrast agent 
for radiology 
in the years 1930-1960 (see for example \cite{Abbatt79}, and the recent 
presentation \cite{Fukumoto12}). 
This 
solution contained radioactive thorium dioxide (ThO$_2$).  
After the first case report\cite{first_case} in 1947 where the 
development of a malignancy was associated with the use of 
Thorotrast 12 years earlier, an endless number of individual case reports 
accumulated evidence on the long-term potential dangers of this substance. 
Thorotrast-related cancers have been reported until recently, 
including one\cite{last_case} where the development of a cancer  
has been associated with the use of Thorotrast 62 years earlier, 
and a recently examined Japanese group whose 
average latent period was 37 years\cite{Yamamoto2}.  

Epidemiological 
investigations extracting disease-specific and total 
risk-factors associated with the use of Thorotrast have been carried 
on and frequently updated 
in several countries, with especially large cohorts in 
Portugal\cite{Portugal03}, 
Denmark \cite{Denmark93}, 
Sweden\cite{Sweden02}, 
Japan\cite{Japan99}, Germany\cite{Germany2008}.  
These studies pose the problem of the 
distribution of the incubation periods over a life-long time scale. 

Latency times are longer than the time needed for the 
growth of the involved cancers, and are 
not related with the time needed for 
thorium dioxide to travel inside the body. 
Since the earlier 
laboratory experimentation (see \cite{SchaeferGolden})  
it has been evident that ThO$_2$ deposits are formed in a short time in those 
same organs where decades later pathologies develop. These deposits 
persist over a time scale that is much longer that a human life 
(see for example \cite{Strole81}). In particular 90 \% of the total 
thorium dioxide burden of the body 
accumulates in liver and spleen (see \cite{KaulNoffz78}), in 
the form of compact granular aggregates large enough to absorb a relevant 
fraction of the emitted $\alpha$ radiation    
(\cite{SchaeferGolden}). 

From the health point of view, in the case of an internal source   
$\alpha$ radiation is especially relevant because of its high energy 
release per unit of path (see for example ch.6 in \cite{Friedlander}). 
Since we are interested in an internal source in close contact with its 
target, $\alpha$ activity will be the object of this analysis. 
The $\alpha$ particles emitted by the decays in the thorium series 
present two features that may be associated with decade-long latency 
periods. 

First, the activity concentration of a sample containing thorium is 
subject to long-term variations until secular equilibrium is  
reached among all the daughter elements of $^{232}$Th (Table 1). 
In a previous work\cite{AB_PM12} the long-term behavior of the 
source activity has been simulated, showing that these variations 
have quantitative relevance 
in the first 10 years after the chemical separation of 
the thorium isotopes from the other elements of a mined ore (this 
process was part of the manufacture of Thorotrast, 
see \cite{Vienna65_1}). 
This behavior depends on the place where the source is present 
inside 
the human body, because tissue-specific biological processes 
may remove some of the daughter elements from the source. 
In the case of Thorotrast injection, 
90 \% of the thorium body burden is formed in liver and spleen. Here  
the activity is decreasing in the first 5-10 years 
and constant later \cite{AB_PM12}. So 
the systematic time-dependence of the activity does not 
explain a long latency period, at least in these organs.  

The thorium series is also characterized by peculiar short-term  
fluctuations in the frequency of emitted particles. 
As evident from Table 1 the decay of $^{228}$Th triggers a series of 
5 $\alpha$ decays in a rather short time. Two of these  
are practically simultaneous, in  
a minute we may have 3 decays, in a day 4 decays, in a  
week 5 decays. In stationary conditions, in a ThO$_2$ source 
whose volume is 1 $\mu$m$^3$ we expect one $^{228}$Th decay per year.  
So a thorium sample has an 
exceptionally small activity in the average, but this activity may 
present dangerous fluctuations in time and space. 
Some authors (\cite{Yamamoto2,Yamamoto2009})  have developed 
models for the onset of the Thorotrast pathologies where an important 
role is played by fluctuations in the absorbed dose at special 
space/time scales. The numerical study of the fluctuations in 
the dose that is absorbed by a small region of tissue 
that is placed near a compact ThO$_2$ source with size $\gtrsim$ 20 $\mu$m 
is the aim of the present work. 

Here an important reference length is the range of the emitted $\alpha$ 
particles. 
Although some emitted particles may have energies up to 8.8 MeV, most 
of them have energies 4-6 MeV, corresponding to ranges of 
15-25 $\mu$m in ThO$_2$ and 40-60 $\mu$m in tissues (see Section 2.2 for  
details). 

Let us assume that a source has a compact structure and a compact shape  
(meaning that its  
density is homogeneous and that its size has the same magnitude in  
different directions) and that a well-defined and regular source-tissue 
separating surface exists. 

We name ``$boundary$'' the source-tissue 
separating surface, ``$distance$'' the distance of a tissue point to 
the source boundary, and $depth$ the distance of a source point to the 
source boundary. 

Intuitively, in the case of a compact source the most exposed 
tissues are those whose distance is much smaller 
than the $\alpha$ particle range. In the worst case (a 
tissue portion that is in touch with the source boundary) the target 
tissue is rarely reached by particles emitted from source points whose
depth overcomes 20 $\mu$m. 
For these reasons, we may guess that the  
radiation reaching a 1-$\mu$m-sized portion of tissue placed within 
a distance of 10 $\mu$m 
is approximately the same 
whether the source size is 20 $\mu$m or 200 $\mu$m. 
So, in the following 
we will speak of a ``large'' source when its size is $>$ 20 $\mu$m, 
and realize a simulation for the limit of an $infinitely$ $large$ source. 
This will reproduce with good precision the radiation intensity 
within 5 $\mu$m of the boundary of a source 
with size 50 $\mu$m, with reasonable approximation the radiation intensity 
within 5 $\mu$m of the boundary of a source with size 20 $\mu$m, and only 
qualitatively the radiation intensity within 5 $\mu$m of a source with 
size 10 $\mu$m. 


We assume a flat and infinite source-tissue separating surface. 
The source fills all the space on one side of this surface, and the 
target all the space on the other side. Such a source may be considered 
the limit of a spherical source with infinite radius. 

We will numerically simulate the process of 
$\alpha$-particle generation and propagation 
for 30 years. 
The two objectives of this work are (i) determine the 
way the average absorbed dose depends on the distance, 
(ii) explore the fluctuations of this radiation absorption during  
intervals of time that are very short with respect to the duration 
of the simulation (from 15 minutes to a week).  
The reference unit of volume to quantify the way the absorbed 
dose is distributed in space will be 1 $\mu$m$^3$ (``detector volume''). 

The numerical code used here has a space precision of 0.1-0.5 $\mu$m, 
and a time precision of 15 minutes. As later discussed (Section 3.2) 
for volumes $>>$ 1 $\mu$m$^3$ or time intervals $\gtrsim$ 1 week the 
frequency 
of many-hit fluctuations may be estimated using Poisson statistics. 
In the region of time and space scales comprised between the code precision 
limits and the limits of validity of the Poisson estimate 
the correlations between serial decays of 
the same nucleus in the thorium series may lead to fluctuations whose 
frequency overcomes much the Poisson estimate 
(see \cite{AB_PM12}). 
For this reason a numerical simulation will be used 
to explore fluctuations in this region. 

Assuming that in a liver a given mass of ThO$_2$ is present, this mass 
may be distributed according to a more compact or sparse geometry. 
If it is 
concentrated in a single compact block (like a sphere) the amount of 
radiation self-absorption by the source is the maximum 
possible. Correspondingly the overall 
dose released to the liver is the minimum possible. If the same ThO$_2$ 
mass is diluted in homogeneously distributed micro-fragments (i.e. a very 
sparse source), self-absorption decreases and the overall dose 
released to the liver is larger. 
This situation is different when the local concentration of radiation 
absorption is considered. In the case of a very sparse source the 
locally absorbed dose is distributed in a roughly homogeneous way. In 
the case of a compact source the local dose absorption 
is dangerously concentrated in the tissues near the source boundary 
(see Section 3.1 for more precise details on this point). For 
this reason this work has been focused on the case of a compact 
source.

\section{Methods} 


The simulation code must  

1) Produce a correct random sequence of nuclear decays 
$D_n(t,x,y,z,E_\alpha,\vec u_\alpha)$. 
The $n-$th nuclear decay $D_n(...)$ takes place in 
the position $x,y,z$ at the time $t$. If the decay product is an $\alpha$ 
particle, it has initial energy $E_\alpha$ and direction given by the 
unitary vector $\vec u_\alpha$. 

2) On the basis of the previous parameters, decide whether this $\alpha$ 
particle may reach a detector. If the answer is positive, the code must 
simulate the trajectory of this $\alpha$ particle, making its energy 
decrease from $E$ $=$ $E_\alpha$ to $E$ $=$ 0, according to the 
local stopping power. 

3) When the $\alpha$ particle crosses a detector, 
record this event  
and evaluate the energy released inside the detector. 

\subsection{Nuclear decays and in vivo evolution of the source composition.}

Thorotrast was made from natural thorium. This excludes the presence 
of many thorium isotopes that are second-generation products of the 
nuclear industry. ``Natural thorium''\footnote{see for example  
\cite{Th_radiochemistry} for general information 
on this and the following points, and \cite{AB_PM12} for a detailed 
study of the evolution of the activity of a thorium 
source. The data in Table 1 have been extracted 
from \cite{ThoriumSeries1}).
}  
is actually a mixture of $^{232}$Th with its daughter elements  
in secular equilibrium (see Table 1 for a list of these elements 
and of some features of their decays). 
Although the daughter elements are less than 1 ppb in weight, each step 
in the chain of decays of Table 1 contributes by the same amount 
(approximately 4000 Bq per gram of thorium) to the 
total activity concentration 40,000 Bq/g, 
whose $\alpha$ component 
is 24,000 Bq/g. 
After mining, a chemical separation 
process is carried on to remove all the non-thorium elements. 
As a result, a mixture of 
$^{232}$Th and $^{228}$Th with relative concentrations 10$^9$:1 
but equal contributions to the total activity concentration 8000 Bq/g 
is available. 

This situation is the starting point of the simulation, corresponding 
to $t$ $=$ 0. 

As studied in \cite{AB_PM12} the 228-component 
is important in the first years after the separation process. 
In a few weeks the decays of $^{228}$Th  
produce all its daughter elements, raising the $\alpha$ activity 
concentration of the sample from 8000 to 24,000 Bq/g.  
After a few years however the initial population of $^{228}$Th and 
of its daughters has disappeared, the activity is much lower, 
and the following evolution is the same as with a starting sample 
of pure $^{232}$Th, reaching again $\alpha$ activity 24,000 Bq/g 
in 10-20 years. This however is a theoretical scenario, with 
relevant changes for in vivo sources. 

Each $\alpha/\beta$ emission is accompanied by the recoil of 
the daughter nucleus. This has two consequences: 

(i) Immediately the recoiling nucleus is displaced over a 
sub-micrometer distance. The recoil energy in this case 
is at most 0.13 MeV (0.05 keV/a.m.u.) corresponding to a final 
range $<<$ 1 $\mu$m (see for example 
the tables on heavy ion ranges in \cite{KayeAndLaby}). This 
displacement may lead to expelling the nucleus from the source. 

(ii) On a much longer time  
scale we may have a slow migration of the nucleus within the source, 
since some of the nuclei of the thorium series do not bind 
themselves to the source 
environment. Even this phenomenon may lead a nucleus to exit the 
source. 

Nuclei that have been expelled from the source are removed from 
the body or dislocated to other organs by biological processes. We will 
generically speak of ``biological depletion'' of the source for all 
the processes of this class. The main effect of biological depletion is 
that in secular equilibrium the daughter/parent activity ratio is 
$<$ 1 for an in vivo source. For some species these ratios 
have been measured and they are reported in the last column of 
Table 2. More precisely the ratio of the activity of a species in a 
tissue sample with respect 
to the activity of $^{232}$Th in the same sample is reported. These 
coefficients are 
taken from \cite{KaulNoffz78}, who refined and summarized the 
results of previous works, and refer to liver and spleen tissues.   

The simulation code does not consider displacements   
of a nucleus inside the source. It however takes into account the possibility 
that these displacements make this nucleus abandon the source, so that 
a probability of biological dislocation of a nucleus from 
the source is included at each step of a decay chain. 




A basic structure in the simulation is a decay chain, that is the series of 
decays of an individual nucleus, starting from the decay 
of a $^{232}$Th nucleus or of a $^{228}$Th nucleus. The first decay of 
a chain is assumed to be randomly  
distributed in the space, with the same probability for each point 
$(x,y,z)$ in all the region occupied by the source material ThO$_2$. 

Because of the very long average life of $^{232}$Th, the 
activity of this nuclear 
species is constant over 30 years. The decay of this nucleus marks the 
beginning of a chain and is an event that is 
homogeneously distributed in time. 
To implement this, when a $^{232}$Th nucleus decays at the time $t_{previous}$, 
a random time $t_1$ is generated with distribution 
$exp(-t_1/T_{no\ decay})$, where $T_{no\ decay}$ is the average time 
separating two independent decays of $^{232}$Th nuclei. Then, the next 
chain starts at the time $t_{next}$ $=$ $t_{previous}\ +\ t_1$. 
With this mechanism, the total number of $^{232}$Th decays from a 
given sample in any assigned time interval 
is a random variable whose long-term average reproduces the 
known $^{232}$Th activity concentration.

\begin{center}
\begin{tabular}
{|lllr|}
\hline
\multicolumn{4}{|c|}{\bf Table 1: nuclear decay data} \\
\hline
parent species \/ & half-life \/ & $\alpha$ energy (MeV) \/ 
& $\alpha$ branching ratio \\ 
\hline
$^{232}$Th & $1.406\cdot 10^{10}$ years & 4.013 & 77 \% \\
  & & 3.954 & 23 \% \\
$^{228}$Ra & 5.739 years & $\beta$ & \\
$^{228}$Ac & 6.139 hours & $\beta$ & \\
$^{228}$Th & 1.913 years & 5.423 & 71 \% \\
  & & 5.340 & 29 \% \\
$^{224}$Ra & 3.632 days & 5.685 & 95 \% \\
  & & 5.440 & 5 \% \\
$^{220}$Rn & 0.9267 min & 6.288 & 100 \% \\
$^{216}$Po & 0.145 s & 6.778 & 100 \% \\
$^{212}$Pb & 10.64 hours & $\beta$ & \\
\hline
branch 1, 35.94 \% & & & \\ 
\hline
$^{212}$Bi & 169 min & 6.090 & 9.6 \% \\
  & & 6.051 & 25.2 \% \\
$^{208}$Tl & 4.4 min & $\beta$ & \\
$^{208}$Pb & stable & & \\
\hline
branch 2, 64.06 \% & & & \\
\hline
$^{212}$Bi & 94.5 min & $\beta$ & \\
$^{212}$Po & $\approx$ 0 & 8.784 & 64.06 \% \\
$^{208}$Pb & stable & & \\
\hline
\end{tabular} 
\end{center}


Once a $^{232}$Th nucleus has decayed producing a $^{228}$Ra nucleus, a random 
time $t_2$ is generated with distribution $exp(-t_2/T_{Ra228})$, where 
$T_{Ra228}$ is the average life of $^{228}$Ra, that is  
the half-life of $^{228}$Ra divided by $ln(2)$. Another number $t_2'$ 
is generated with distribution $exp(-t_2'/T_{Ra228.BIO})$, where 
$T_{Ra228.BIO}$ is the average time required for the biological 
removal of $^{228}$Ra from this tissue. The times $t_2$ and $t_2'$ 
are compared, and if $t_2$ $<$ $t_2'$ the nuclear decay is ``accepted'' 
and the chain of decays may continue. If $t_2$ $>$ $t_2'$ the nucleus has 
been biologically removed from the source before its decay could take 
place, and this chain stops here.

\begin{center}
\begin{tabular}
{|l|l|l|l|}
\hline
\multicolumn{4}{|c|}{\bf Table 2: biological depletion data} \\
\hline
parent \/ & physical \/ & bio/physical \/ 
& final activity \\ 
species \/ & half-life \/ & half-life ratio\/ 
& ratio to $^{232}$Th\\ 
\hline
$^{228}$Ra & 5.739 years & 0.665 & 0.38\\
$^{228}$Th & 1.913 years & 7.2 & 0.35 \\
$^{224}$Ra & 3.632 days & 2.25 & 0.24 \\
$^{220}$Rn & 0.9267 min & rem.prob. 20\% & 0.19 \\
$^{212}$Pb & 10.64 hours & 1.1 & 0.10 \\
\hline
\end{tabular} 
\end{center}





If the chain continues, the later decays in the chain -  
of $^{228}$Ra, $^{228}$Ac, 
$^{228}$Th, $^{224}$Ra, $^{212}$Pb, $^{212}$Bi - 
are handled the same way.
The other radionuclides are too short-lived to make their actual 
lifetimes relevant for the temporal resolution, limited to 15 minutes 
in the current analysis for the following reasons. In single precision (six 
digits) the code is not able to distinguish time scales under 0.01 days 
if we stop at the midnight of the 9999th day (after this day the 
allowed time resolution is 0.1 days). All the intermediate steps 
in the calculation require expressing times in double precision and this 
is done, but the final results are recorded in single-precision lists 
nonetheless, 
so the date recording is precise within 0.01 days (15 minutes) until the 
9999th day, and 0.1 days thereafter.

Within a time resolution 0.01 days 
it is correct to consider the three decays of 
$^{224}$Ra, $^{220}$Rn, and $^{216}$Po as simultaneous, since the probability 
of having more than 15 minutes between the first and the last has magnitude 
$10^{-5}$ (this may introduce some error in the results of 
fig.\ref{fig:fluctuation0_15}, but a small error in any case).  
For the same reasons also the 
decays following that of $^{212}$Bi are treated as simultaneous. 

This also poses the problem of implementing the 
biological removal of $^{220}$Ra, 
since in this case both the nuclear and the biological half-lives are 
below the time precision scale, but their competition is relevant. 
In this case, the code just chooses whether 
this nucleus has been removed according to a fixed probability reported 
in Table 2.  

At the beginning of the process the source contains a  
$^{228}$Th contamination able to produce exactly the same activity as 
$^{232}$Th. To simulate this initial condition, 200 years of $^{232}$Th 
decays have been simulated $before$ the starting time $t$ $=$ 0, 
without track reconstruction. These 
decays, each followed by a chain of decays of the daughter nuclei, 
produce a correct secular equilibrium population 
of $^{228}$Th nuclei at $t$ $=$ 0. The non-thorium nuclei produced by these 
``before-zero'' decays are removed at $t$ $=$ 0, as would 
happen in a chemical filtering process. 
Those $^{228}$Th nuclei that are 
present in the source at $t$ $=$ 0 may begin a series of decays and 
$\alpha$ particle productions, exactly as $^{232}$Th nuclei do. 

When a nuclear decay is an $\alpha$-decay  
there is more than one possibility, as shown in Table 1. 
For example, in the case of $^{224}$Ra, 
an $\alpha$ with energy 5.685 MeV is emitted with probability 95 \%, while 
an $\alpha$ with energy 5.440 MeV is emitted with probability 5 \%. 
The code randomly selects one of the two with a relative probability 
of 95:5.

When $^{212}$Bi decays the chain may follow two different paths (see Table 1). 
The code 
randomly chooses one of the two with the associated probabilities. 
The probability of the first path is 35.94 \% and that of the second 
64.06 \%. 

Although the time required to the $\alpha$ particle to lose all its 
initial energy is finite and measurable, it is considered zero here, 
since it is many orders smaller than the time resolution 0.01 days used 
here. This is a relevant simplification, since it means that 
the code is not required to handle two tracks simultaneously. 


The parameters of the biological depletion in the last column of 
Table 2 are the long-term relative activities of some nuclear 
species with respect to the one of 
$^{232}$Th, as reported in \cite{KaulNoffz78}. These numbers 
are not directly used in the code. Rather, the code needs the biological 
half-lives (or the probability of removal for the case of $^{220}$Rn). 
So a first relevant step in the present work has been to calculate 
these parameters.  

This has been done by a series of attempts. The simulation starts with pure 
$^{232}$Th, and we insert a set of biological half-lives in the code, adjusting 
them until a long-term simulation reproduces the activity ratios of 
Table 2. These do not depend on 
the initial composition of the source \cite{AB_PM12}, 
so the simple choice of ``pure $^{232}$Th at $t$ $=$ 0'' is sufficient for 
this analysis. The values of the biological half-lives that allow the 
asymptotic ratios to coincide with the measured values are reported 
in Table 2. 

\subsection{Structure and properties of the source and of the 
target/detector regions.} 

Once an $\alpha$ particle has been produced with a random orientation for 
its velocity, the code must simulate its trajectory and energy release 
in matter. 
Two materials are crossed by an $\alpha$ particle, each with its 
stopping power features (for the stopping powers and ranges needed here, see 
\cite{NIST,SRIM}):  

a) The source material ThO$_2$. The stopping 
power of this molecular material has been chosen as 
a linear combination of the composing 
elements, each weighted by its mass fraction in the molecule 
(Bragg's rule\cite{Bragg}). 

b) The organic tissue, presently approximated by liquid water. 
Stopping power features of muscular tissue and other tissues 
are an easily available alternative but they are very similar to 
those of water (see data tables and graphs at the NIST site\cite{NIST}). 

\begin{figure}[ht]
\centering
\includegraphics[width=9cm]{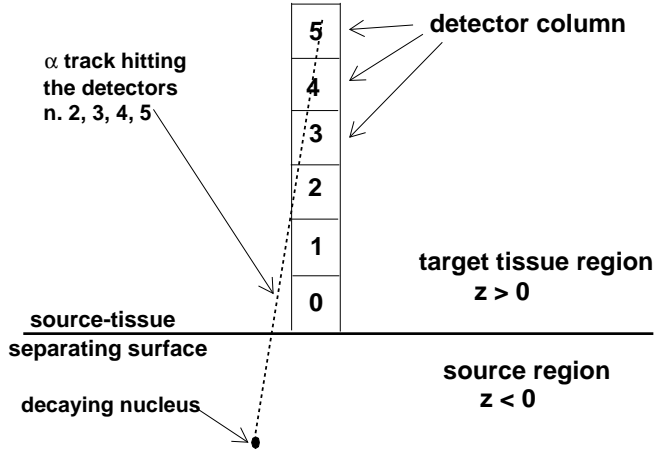}
\caption{
Structure of the source-target-detector region. Referring to cartesian 
coordinates $x,y,z$, the $xy$-plane is the 
source-target separating surface (``source boundary'' in the text). 
The source fills the lower 
half-space ($z$ $<$ 0), while the upper half-space ($z$ $>$ 0) 
is filled with the target tissue. In the target region we see the 
detector column (whose central axis coincides with the $z$-axis). 
This is a stack of 115 independent detectors, six of which are visible 
in the figure. Each detector is a 1 $\mu$m$^3$ cube, with center in 
position $(0, 0, 0.5)$, $(0, 0, 1.5)$, etc.  
\label{fig:geometry}} 
\end{figure}


The range of energies of the $\alpha$ particle of interest here is 
1 keV to 9 MeV. 
In water the maximum energy loss is 2266 MeV cm$^2$/g (i.e. 0.227 MeV/$\mu$m) 
at $E$ $=$ 0.7$-$0.75 MeV. This value is the peak 
energy loss on the left of the minimum of the Bethe-Bloch
curve (see for example \cite{Friedlander}). 
The minimum itself would be at much higher energies than 9 MeV. 
In the range 0.001-9 MeV  
the distance needed to release a given amount 
of energy is 2.5-3 times larger in water than in ThO$_2$. 
With 10 MeV of initial energy, an $\alpha$ track cannot cover more
than 40 $\mu$m in ThO$_2$ and 115 $\mu$m in water. These range 
values have been used to determine the size of the implemented 
source and of the detector column. 
The modeled source is infinitely extended, but 
in its practical 
implementation it is useless to consider source regions from which an 
emitted particle could never reach the detector. 

The physical model is idealized by the 3-dimensional 
space of points $(x,y,z)$ 
divided into two regions (see fig.\ref{fig:geometry}): 
for $z$ $<$ 0 we have ThO$_2$ and for 
$z$ $>$ 0 water.  

The implemented source region is a volume 
$\Delta x \Delta y \Delta z$, with $\Delta x$ $=$ $\Delta y$ $=$ 
230 $\mu$m, and $\Delta z$ $=$ 40 $\mu$m. More precisely, 
the ThO$_2$ source region is: 

$-115$ $\mu$m $<$ $x$ $<$ $+115$ $\mu$m, 

$-115$ $\mu$m $<$ $y$ $<$ $+115$ $\mu$m, 

$-40$ $\mu$m $<$ $x$ $<$ 0, 

The detector column covers 115 $\mu$m 
of $z$-axis (see fig.\ref{fig:geometry}): 

$-0.5$ $\mu$m $<$ $x$ $<$ $+0.5$ $\mu$m, 

$-0.5$ $\mu$m $<$ $y$ $<$ $+0.5$ $\mu$m, 

0 $<$ $x$ $<$ 115 $\mu$m. 

It is divided into 115 cubes, each with volume equal to 1 $\mu$m$^3$.  
Each of them is an independent detector. This point needs 
to be stressed: the basic detector is a cube, not the column. 
The column is a 
collection of 115 independent detectors, that allow one to use one and 
the same simulation to evaluate what happens in detectors put 
at different distances to the source boundary. 



Both in the target and in the source regions  
the energy loss by an $\alpha$ particle during its travel 
is computed to update the particle energy at each step of the 
simulated trajectory. 
Inside the detector volume, the energy released in each 
cubic sub-region and the time of the particle passage are recorded. 
The steps are short enough to have many steps inside each detector 
volume. 

\subsection{Straight trajectories and selection rules.}

The reconstruction of the tracks of all the generated particles for 
30 years of simulated time would require too much CPU on  
an ordinary computer, taking also into account that many 
simulations are needed for fine tuning the code and accumulating 
statistics. Within some approximations it is possible 
to reduce much the number of particles whose trajectories 
are reconstructed by the code. 

A key approximation used in this code is that a trajectory is 
straight. This allows for deciding from the very 
beginning whether a track will reach the detector or will not. 

Generated $\alpha$ particles may be discarded from the very beginning 
based on the following two decisions made by the algorithm: 

a) A particle has insufficient energy to reach any portion of the detector. 

b) The straight-line extrapolation of the particle trajectory will not 
cross any detector. 

The latter warrants some discussion, because it limits the space 
precision of the code to 0.1-0.5 $\mu$m for the final parts 
of a reconstructed track. 

The quoted databases for the stopping power of the $\alpha$ particles 
in matter give stopping power and ranges both relative to the projected 
path (i.e. assuming the same hypothesis of straight-line trajectories 
used here) and to the length of the real path. A comparison of projected 
and real ranges \cite{NIST} of $\alpha$ particles 
in water shows that their difference 
is about 0.1 $\mu$m at 10 keV compared to 
a residual range of 0.3 $\mu$m, and 
0.2 $\mu$m at 60 keV compared to a residual 
range of 1 $\mu$m. For larger energies both the true and the projected range  
increase, but their difference is stable, implying that this difference 
is built in the last 10-60 keV of energy and that  
the last 0.3-1 micrometers 
of trajectory deviate from a straight line. 

If this were due to a single deflection at 60 keV, triangle geometry would 
mean a deviation from the straight line equivalent to 40$^o$, 
that is a transverse shift of 0.6 $\mu$m. This is an extreme case,  
a standard deflection is considerably smaller. So we estimate an error 
of 0.1-0.5 $\mu$m in the position where the particle comes 
to rest.


\section{Results} 

\subsection{Systematic trends.}

\begin{figure}[ht]
\centering
\includegraphics[width=9cm]{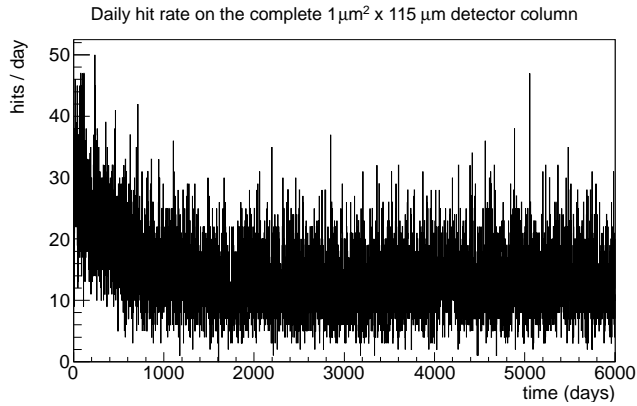}
\caption{
Daily hit rate for the full detector column (a stack of 115 detectors), 
simulated over a 30 year period. 
\label{fig:full_column_rate}}
\end{figure}

Fig.\ref{fig:full_column_rate} shows the number of $\alpha$ particles  
that strike the full detector column daily over 30 years. 
It is possible 
to distinguish the average trend and its fluctuations. The 
time needed to reach a stationary dose rate 
(daily fluctuations apart) is about 5 years, that is 
shorter than the time needed to a thorium source in nature to  
reach secular equilibrium (a few decades). As noted in 
\cite{AB_PM12} this is a consequence of biological 
depletion. 

\begin{figure}[ht]
\centering
\includegraphics[width=9cm]{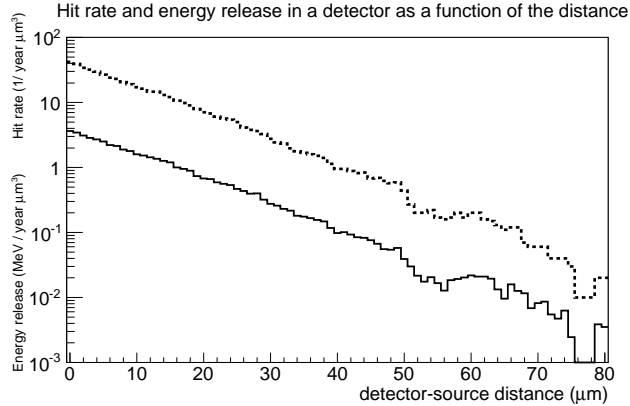}
\caption{
Energy release and hit frequency in a detector (a 1 $\mu$m$^3$ cube) 
as a function of the 
distance between the detector and the source boundary. 
Continuous histogram: Energy released in each detector 
in 1 year in stationary conditions (20 years 
after the beginning of the simulation). Short-dashed histogram: 
number of hits in each detector 
in 1 year. 
\label{fig:dose_distance}}
\end{figure}

Fig.\ref{fig:dose_distance} shows how these hits and the associated 
energy releases are distributed with respect to the 
distance to the source-tissue separating surface. 
The figure reports the cumulative 
number of hits in one year on each detector of the column, 
and the total amount of energy released in each detector. In 
the region within 70 $\mu$m from the source the 
two histograms are almost exactly proportional, with an average 
0.09 $\pm$ 0.005 MeV$/\mu$m$^3$  
of released energy for each hit.  
The decrease of the hit frequency 
and the absorbed energy with the distance are almost exactly 
exponential, and one order of magnitude is lost in 25 $\mu$m.  

This suggests that, as far as the time-averaged dose absorption 
is concerned, in presence of a large and compact source 
the really exposed region of the target is a layer adjacent to 
the source with thickness of 25 $\mu$m.

\subsection{Fluctuation analysis - general criteria of analysis and relevant space-time scales.}


When speaking of ``$n$ hits in 24 hours'' (in a specified detector) 
one may refer to two classes of event: 

1) The total simulated time (31 years) is divided into 
1-day-long bins. Each day begins and ends at midnight and $n$ is the 
number of hits recorded during that day. 

2) A series of $n$ hits may end within 24 hours from its beginning, but 
these $n$ hits do not necessarily take place in the same day. In this 
case we may have partially overlapping series of hits. 

The former case is easier to handle for statistics. 
The latter is more relevant for health or safety considerations.
Results will be presented for both. The following results are averages 
from a set of 10 simulations 
(for the analysis of the events of the 
class (1)) or 30 simulations (for the analysis of the events of the 
class (2)). So it is not strange to have results like 
``0.3 events''. Each simulation is 31-year long, with the exception of 
the 15-minute-binning case where each simulation covers 
10,000 days. 

We have to do with three variables, that we name: 

``Number of hits'' the number of particles incident on a detector in a 
given time interval.  

``Volume'': the volume of the detector where the number of hits is recorded. 
In this preliminary analysis this is not necessarily 1 $\mu$m$^3$. 

``Time interval'': the time interval during which the hits are 
regrouped together to form a fluctuation, for example 1 day or 1 week. 
This must not be confused with the much longer duration of the simulation. 

We also define ``time scale'' and ``volume scale'' the magnitude of the 
time interval and of the volume. 

The number of possible combinations of hit number, time scale and 
volume scale is rather broad. A numerical simulation should primarily 
be focused on those combinations for which alternative 
analytical methods are not well justified. So a relevant preliminary problem 
is to understand to what extent one may rely upon 
the Poisson distribution for estimating fluctuations in the number 
of hits within a given time interval and volume. 

Poisson statistics is inappropriate when a relevant fraction of 
the $n$ hits recorded in 
a time interval is constituted by serial decays 
of one and the same nucleus. An examination 
of the thorium series shows that we have three critical emission 
frequencies: 

A) 3 decays in a minute, from the sequence of decays 
$^{224}$Ra $\rightarrow$ 
$^{220}$Rn $\rightarrow$ 
$^{216}$Po $\rightarrow$ 
$^{212}$Pb. 

B) 4 decays in a day, from the sequence 
$^{224}$Ra $\rightarrow$ 
$^{220}$Rn $\rightarrow$ 
$^{216}$Po $\rightarrow$ 
$^{212}$Pb $\rightarrow$ 
$^{208}$Pb.  

C) 5 decays in a a week, from the sequence of decays 
$^{228}$Th $\rightarrow$ 
$^{224}$Ra $\rightarrow$ 
$^{220}$Rn $\rightarrow$ 
$^{216}$Po $\rightarrow$ 
$^{212}$Pb $\rightarrow$ 
$^{208}$Pb.  

Consider the first case as an example. If a detector is 
hit with an average frequency 4000/day, then we expect a Poisson 
distribution for the number of hits/day, because these 4000 hits 
cannot be all serial decays of one and the same nucleus. 
On the other side, these 4000 hits/day mean 3 hits/minute, and these 
3 hits $might$ come from consecutive decays of a single  
nucleus. The conditions for a safe application 
of the Poisson statistics are $a$ $priori$ present at the time scale 
of 1 day, but not at the time scale of 1 minute. 

To generalize this example, for the thorium series we define the three critical 
hit frequencies: 

$F_A$ $=$ 3/minute $\approx$ 4000/day, 

$F_B$ $=$ 4/day, 

$F_C$ $=$ 5/week $\approx$ 0.7/day. 

with $F_A$ $>>$ $F_B$ $>>$ $F_C$. 

Let $F(V)$ be the average frequency of hits 
on a detector of volume $V$. 

If $F(V)$ $>>$ $F_A$, we expect the Poisson distribution to be a good 
estimator for multiple-hit events at all time scales, because $F(V)$ is much 
larger than all the critical hit frequencies. 

If $F_B$ $<<$ $F(V)$ $\lesssim$ $F_A$,  
groups of 2 or 3 events may escape the 
Poisson expectation on a time scale 1 minute, but for time scales like 1 day 
or 1 week the Poisson distribution is justified. 

If $F_C$ $<<$ $F(V)$ $\lesssim$ $F_B$,  
groups of 2, 3 or 4 events may escape the Poisson expectation on a time scale 
1 day or smaller, while for a time scale of 1 week 
the Poisson expectation is justified.  

If $F(V)$ $\lesssim$ $F_C$, groups of up to 5 events may escape the Poisson 
expectation for time scales up to 1 week. For longer time scales the 
Poisson expectation is justified. 

We have $F(1$ $\mu$m$^3)$ $=$ 40 hits/$\mu$m$^3$year 
$\approx$ 0.1 hits/$\mu$m$^3$day 
for a target region that is in touch with the source 
(see fig.\ref{fig:dose_distance}). This means that for the assigned 
source $F(1 \mu$m$^3)$ $<$ $F_C$ $<$ $F_B$ $<$ $F_A$. 

So, in a detector of volume 1 $\mu$m$^3$ we may expect non-Poisson 
fluctuations on a time scale of 1 week or smaller. 
However, since $F(V)$ is an increasing function of $V$, increasing 
$V$ over 1 $\mu$m$^3$ we will find 
volumes $V_C$, $V_B$, and $V_A$  
such that: 

$F(V_C)$ $=$ $F_C$, 

$F(V_B)$ $=$ $F_B$, 

$F(V_A)$ $=$ $F_A$. 

If we neglect the dependence of the hit frequency on the distance 
and assume $F(V)$ $\propto$ $V$, we have 

$V_C$ $=$ 7 $\mu$m$^3$ (a cube of about 2 $\mu$m of 
edge). Above this volume, the number of events/week follows Poisson 
expectation. 

$V_B$ $=$ 40 $\mu$m$^3$ (a cube of 3-4 $\mu$m of edge). Above 
this volume, the number of events/day follows the Poisson expectation. 

$V_A$ $=$ 40,000 $\mu$m$^3$ (a cube of 0.2 mm of edge). Above this 
volume, the number of events/minute follows the Poisson distribution. 

For these reasons, in the following the reference detector 
volume will be 1 $\mu$m$^3$. Three different time scales between 
the code precision limit 15 minutes and the week will be considered. 



\subsection{Fluctuation analysis - fixed time binning.}

\begin{figure}[ht]
\centering
\includegraphics[width=9cm]{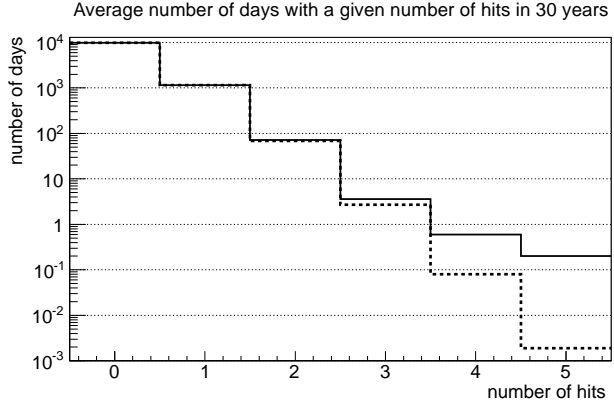}
\caption{
Continuous histogram: Average number of days with a given 
number of hits in the first detector 
in 30 years. The average is 
calculated over 10 repetitions of the simulation. So, 0.1 is the minimum 
possible number of days, resulting from one day only in 10 simulations. 
Short-dashed histogram: the Poisson distribution estimate. 
\label{fig:fluctuation0}}
\end{figure}

Figure \ref{fig:fluctuation0} 
shows the distribution 
of days with $n$ hits and the corresponding Poisson expectation 
in the first detector, the one in touch with the 
source boundary. 

A consideration of similar distributions in detectors at larger 
distances shows that the distribution in the second detector is 
qualitatively similar to the one of 
fig.\ref{fig:fluctuation0} but the number of 4-hit and 5-hit events 
is much smaller (one 5-hit/day event in 10 simulations). 
For larger distances, only 
once we find a day with 4 hits (at distance of 4 $\mu$m). At all 
distances, for the frequencies $\leq$ 3 hits/$\mu$m$^3$day the 
discrepancies between 
the simulated distribution and the Poisson one are not relevant. 

\begin{figure}[ht]
\centering
\includegraphics[width=9cm]{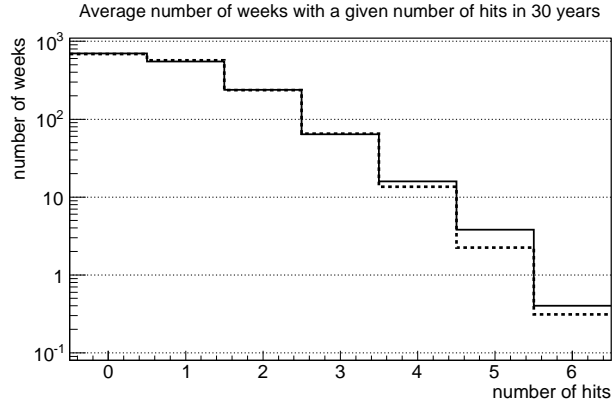}
\caption{
Average number of weeks with a given 
number of hits in the first detector in 30 years. The average is 
calculated over 10 repetitions of the simulation. 
Short-dashed histogram: the Poisson distribution estimate. 
\label{fig:weeks}}
\end{figure}

Fig.\ref{fig:weeks} shows the distribution of $weeks$ with $n$ hits  
in the first detector volume. This 
distribution is close to the Poisson one. The worst 
disagreement is a factor 3/2 in the case of 5-hit events, the same 
value at which we find the largest disagreement between the Poisson 
expectation and the simulated numbers in figure \ref{fig:fluctuation0}. 

\begin{figure}[ht]
\centering
\includegraphics[width=9cm]{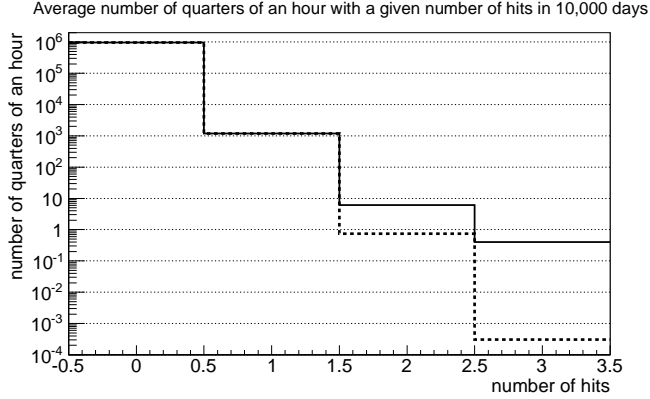}
\caption{
Average number of 15-minute periods with a given 
number of hits in the first detector in 10,000 days (slightly 
more than 27 years). The average is 
calculated over 10 repetitions of the simulation. 
Short-dashed histogram: the Poisson distribution estimate. 
\label{fig:fluctuation0_15}}
\end{figure}

Fig.\ref{fig:fluctuation0_15} 
shows the distribution of the hits in 15-minutes periods. Here the 
number of 2-hit-periods is one order of magnitude larger than the Poisson 
expectation, and the number of 3-hit-periods several orders of 
magnitude larger. 

The numerical simulations partially confirm what suggested by the preliminary 
analysis. For 4-hit and 5-hit events on a target with 
volume equal to 1 $\mu$m$^3$, a day seems a critical time scale, with 
a frequency of such events that largely overcomes the Poisson prediction. 
Expanding the time scale to 1 week, one arrives to the 
border of the critical region: the Poisson 
estimate for such events is almost correct. 

The frequencies of 2-hit and 3-hit events are Poisson-like at the 1-day scale. 
They are larger by orders than the Poisson estimate at the 
shortest scale allowed by this code, 15 minutes. 

\subsection{Fluctuation analysis - frequency of ``within 24 hours'' 
fluctuations.} 

For this analysis 30 independent simulations of 30 years each 
have been performed. In these simulations we have found 7 ``extreme''   
and not reciprocally overlapping fluctuations on the 24-hour and 
1 $\mu$m$^3$ scales:  

\noindent 
One 6-hit event within 24 hours and 15 minutes 
(this involves two overlapping 5-hit series within 24 hours each). 

\noindent 
Six 5-hit events within 24 hours (in addition to the previous two events 
that have been counted as a single 6-hit event). 
 
The above 7 sequences were homogeneously distributed in 30 years.  
The first years (with a relevant weight of $^{228}$Th-initiated chains)   
did not show a special frequency of many-hit events. 

Four of these sequences contained 5 particles emitted 
by the same nucleus, 
one contained 2 particles only by the same nucleus, and 
two sequences contained 5 $\alpha$ particles all emitted by different 
nuclei. Only 
the two sequences with all the particles emitted by different nuclei 
would be predicted by the 
Poisson distribution, so from another point of view we may confirm 
that this statistics is inadequate to predict the frequency of days 
with many hits on the micrometer scale, at small 
distances from the source. 

Only one of the above sequences involved a detector 
detector that was not the first one of the column, 
and it was a case where all the 
hits came from different nuclei. Probably, the relative weight of 
those events where all the incident particles come from one and the same 
nucleus decreases with the distance. 


4-hit events in 24 hours in 1 $\mu$m$^3$ may be found at larger distances. 
The space distribution of such events in the detector 
column in 30 years and 30 simulations is (those overlapping with the 
previous 5-hit and 6-hit events have been excluded): 

28 at distance 0-1 $\mu$m, 10 at 1-2 $\mu$m, 15 at  
2-5 $\mu$m, 9 at 5-10 $\mu$m, 3 at 10-20 $\mu$m 
(the last found event was at 19 $\mu$m). 

Combining all the events 
with 4 $or$ $more$ hits would increase the number of events 
at distance 0-1 $\mu$m to 34, and the number of events at distance 
1-2 $\mu$m to 11. This means that 60 such events out of a total of 72 
are within 5 $\mu$m of the source, and 69 out of 72 within  10 $\mu$m.

\section{Conclusions and perspectives} 

Summarizing the main results: 

The time-averaged frequency of hits and the time-averaged released dose 
decreases exponentially with the distance from the source-tissue 
separating surface, in such a way that an order of magnitude 
in the particle and energy flux is lost in about 25 $\mu$m from the 
source. So, if the development of a cancer 
is associated with single-hit events, the really dangerous region should 
coincide with this 25-$\mu$m-thick 
layer of tissue that is in touch with the source. 


The considered statistics (up to 30 simulations) 
has allowed analyzing the frequency of 
events with up to 5 hits within 24 hours in the same target volume. 
An event with 6 hits have been seen, but just one. 

On target regions of size 1 micrometer, events with 4-hit per day and 5-hit 
per day take place with a frequency that is much larger than  
predicted by the Poisson statistics. The same is true for 
events with 3 hits in a much shorter time interval (15 minutes). 
If the development of a cancer 
is associated with such multi-hit events, the dangerous region is 
is a layer of tissue that is in touch with the source, with a thickness 
that is a decreasing function of the 
number of hits. It may be estimated as 1 $\mu$m for 5-hit events, 5-10 $\mu$m 
for 4-hit events. 

If the time interval is increased to 1 week or more,  
the Poisson statistics gives good predictions for multi-hit events on 
target regions with size of 1 $\mu$m. 
The same is expected for a  
time interval of 1 day if the size of the considered target 
region is $>>$ 1 $\mu$m. Events with 
2-3 hits in a minute are expected to overcome Poisson estimates on any 
target region of sub-visible size. 




Further work on the preliminary analysis presented is, of course, 
required.  Most notably, this includes improving the statistics 
(i.e. the number of simulations) of the analysis, for example, to more 
robustly characterize dependence of many-hit events on distance.  Other 
important refinement would be consideration of different source 
structure and densities and of sparse granular source distributions.

\vspace{1truecm}
\noindent
{\bf Ethical standards:} No human-involving study is reported here.

\vspace{0.5truecm}
\noindent
{\bf Conflict of interest statement:} None declared.



\end{document}